\documentclass[aps,prb,showpacs,twocolumn,superscriptaddress]{revtex4} 

\usepackage[dvipdfmx]{graphicx}
\usepackage[version=3]{mhchem}

\begin{document}

\title{Materials informatics based on evolutionary algorithms: Application to search for superconducting hydrogen compounds}



\author{Takahiro Ishikawa}%
 \email{ISHIKAWA.Takahiro@nims.go.jp}
 \affiliation{%
 ESICMM, National Institute for Materials Science, 1-2-1 Sengen, Tsukuba, Ibaraki 305-0047, Japan
 }%
 \affiliation{%
 Center for Science and Technology under Extreme Conditions, Graduate School of Engineering Science, Osaka University, 1-3 Machikaneyama, Toyonaka, Osaka 560-8531, Japan
 }%
\author{Takashi Miyake}%
 \affiliation{%
 CD-FMat, National Institute of Advanced Industrial Science and Technology, 1-1-1 Umezono, Tsukuba, Ibaraki 305-8568, Japan
 }%
\author{Katsuya Shimizu}%
 \affiliation{%
 Center for Science and Technology under Extreme Conditions, Graduate School of Engineering Science, Osaka University, 1-3 Machikaneyama, Toyonaka, Osaka 560-8531, Japan
 }%

\date{\today}

\begin{abstract}
We present materials informatics approach to search for superconducting hydrogen compounds, which is based on a genetic algorithm and a genetic programming. 
This method consists of four stages: (i) search for stable crystal structures of materials by a genetic algorithm, (ii) collection of physical and chemical property data by first-principles calculations, (iii) development of superconductivity predictor based on the database by a genetic programming, and (iv) discovery of potential candidates by regression analysis. 
By repeatedly performing the process as (i) $\rightarrow$ (ii) $\rightarrow$ (iii) $\rightarrow$ (iv) $\rightarrow$ (i) $\rightarrow$ $\dots$, the superconductivity of the discovered candidates is validated by first-principles calculations, and the database and predictor are further improved, which leads to an efficient search for superconducting materials. 
We applied this method to hypothetical ternary hydrogen compounds and predicted \ce{KScH_{12}} with a modulated hydrogen cage showing the superconducting critical temperature of 122\,K at 300\,GPa and  \ce{GaAsH_{6}} showing 98\,K at 180\,GPa. 
\end{abstract}


\maketitle


\section{Introduction}

The use of an informatics approach to materials science, \textit{i.e.} materials informatics (MI),  has been expected to bring the acceleration for the exploration of new materials~\cite{Potyrailo2011,Jain2013,Ong2013,Takahashi2016}. 
High-throughput screening and machine learning 
have been employed 
to discover hidden trends and create predictive models in databases of 
physical and chemical properties for crystalline compounds, \textit{e.g.} 
search for cathode materials with a long cycle life for lithium-ion battery~\cite{Nishijima2014} and new superconducting materials~\cite{Stanev2018-ML}. 

Search for new physical and chemical properties 
have been performed by varying 
compositions and external parameters such as pressure, temperature, and electromagnetic field with respect to already known materials. 
These approaches are considered as the exploration of optimal solutions in vast and complicated search space under given conditions. 
In such a case, evolutionary algorithm (EA), which is a heuristic-based approach to solving problems using mechanisms inspired by biological evolution, \textit{e.g.} mating, mutation, selection, inheritance, \textit{etc.}, is effective for the discovery of the optimal solutions. 
EA has been used in a wide variety of scientific research fields and has been applied to search for stable or metastable crystal structures~\cite{Deaven1995,Bush1995,Woodley1999,Woodley2004,Oganov06}, 
well-performed mathematical equations and computer programs in predefined task, and so on. 

\begin{figure}
\includegraphics[width=8.2cm]{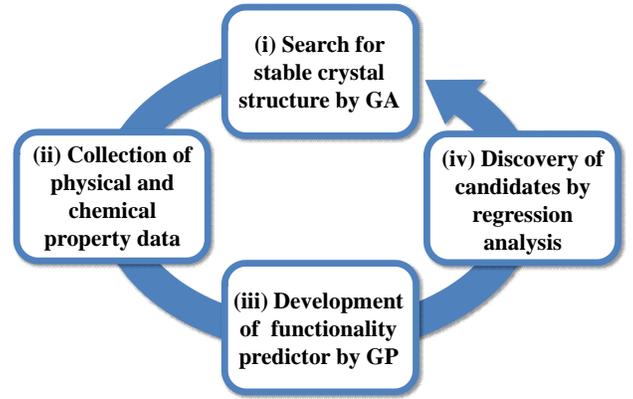}
\caption{\label{Fig-MI} 
(Color online) Materials informatics approach to search for new functional materials, which is based on the genetic algorithm (GA) and the genetic programming (GP).}
\end{figure}
In this study, we propose materials informatics approach to search for new functional materials, which is based on a genetic algorithm (GA) and a genetic programming (GP) included in the EA techniques (Fig. \ref{Fig-MI}). This method consists of four stages: (i) determination of stable crystal structures of materials by GA, (ii) collection of physical and chemical property data by preforming first-principles calculations and development of database, (iii) development of a functionality predictor based on the data by GP, and (iv) discovery of potential candidates by regression analysis. 
By repeatedly performing the process as (i) $\rightarrow$ (ii) $\rightarrow$ (iii) $\rightarrow$ (iv) $\rightarrow$ (i) $\rightarrow$ $\dots$, 
the candidates discovered at the stage (iv) are validated by first-principles calculations through (i) to (ii), and the database and predictor are gradually improved using the validation results. Thus, this method enables to accelerate the search for new functional materials. 
This paper is organized as follows: 
the details of GA for crystal structure search and GP for predictor development are presented in Sec. \ref{EA}, 
the application of our MI method to the search for superconducting hydrogen compounds is shown in Sec. \ref{Application}, and the discussion and conclusion are drawn in Sec. \ref{CON}. 

\section{\label{EA}Evolutionary algorithms}
 
\begin{figure}
\includegraphics[width=8.5cm]{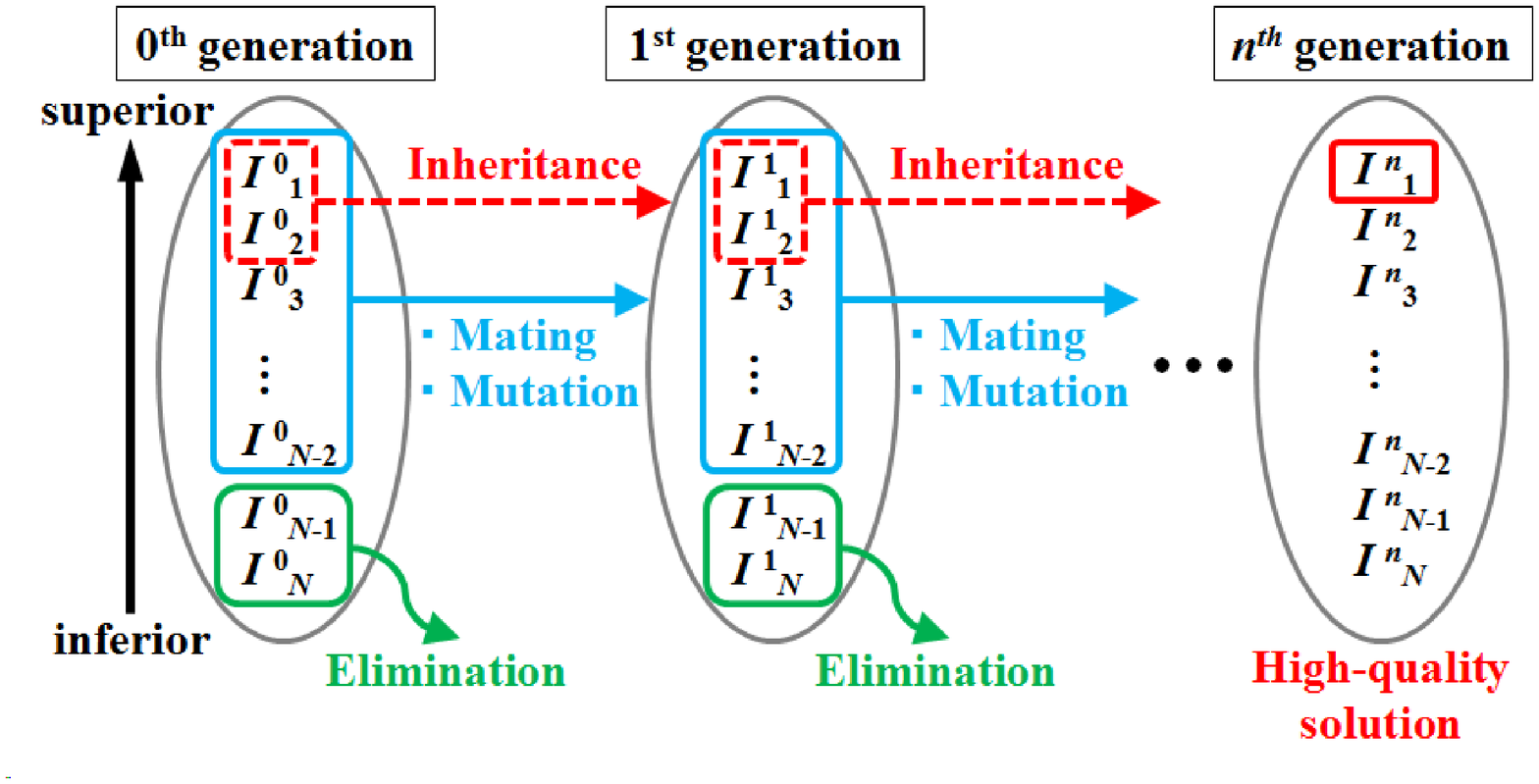}
\caption{\label{Fig-EA} 
(Color online) Schematic view of evolutionary algorithm. $I$ represents an individual, and $N$ is the number of individuals in the population. }
\end{figure}

Figure \ref{Fig-EA} shows a process on the search for high-quality optimal solutions 
by EA. First, we prepare for a population consisting of individuals ($I$), which are generated randomly and ranked according to a fitness parameter (0th generation). 
Then, few especially inferior individuals are eliminated from the population (elimination), and the other ones are used for the creation of new individuals for the next generation.  The new individuals are created by 
randomly applying evolutionary operations, ``mating'' and ``mutation''. 
Few especially superior individuals are inherited to the next generation (inheritance), and 
all the individuals are ranked again.  
By repeatedly performing this process, the population is evolved to more superior one and finally high-quality optimal solutions are obtained. 
As mentioned above, in our MI approach, we use GA to search for stable crystal structures at the stage (i), and GP to develop a functionality predictor at the stage (iii).  

\subsection{GA for crystal structure search}

\begin{figure}
\includegraphics[width=8.5cm]{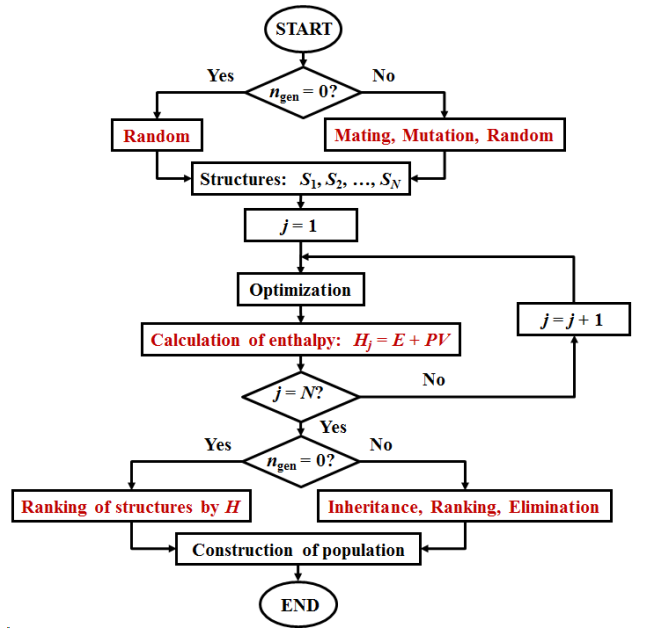}
\caption{\label{Fig-flowchart-GA} 
(Color online) Flowchart at each generation with respect to crystal structure prediction by GA.}
\end{figure}

\begin{figure}
\includegraphics[width=8.5cm]{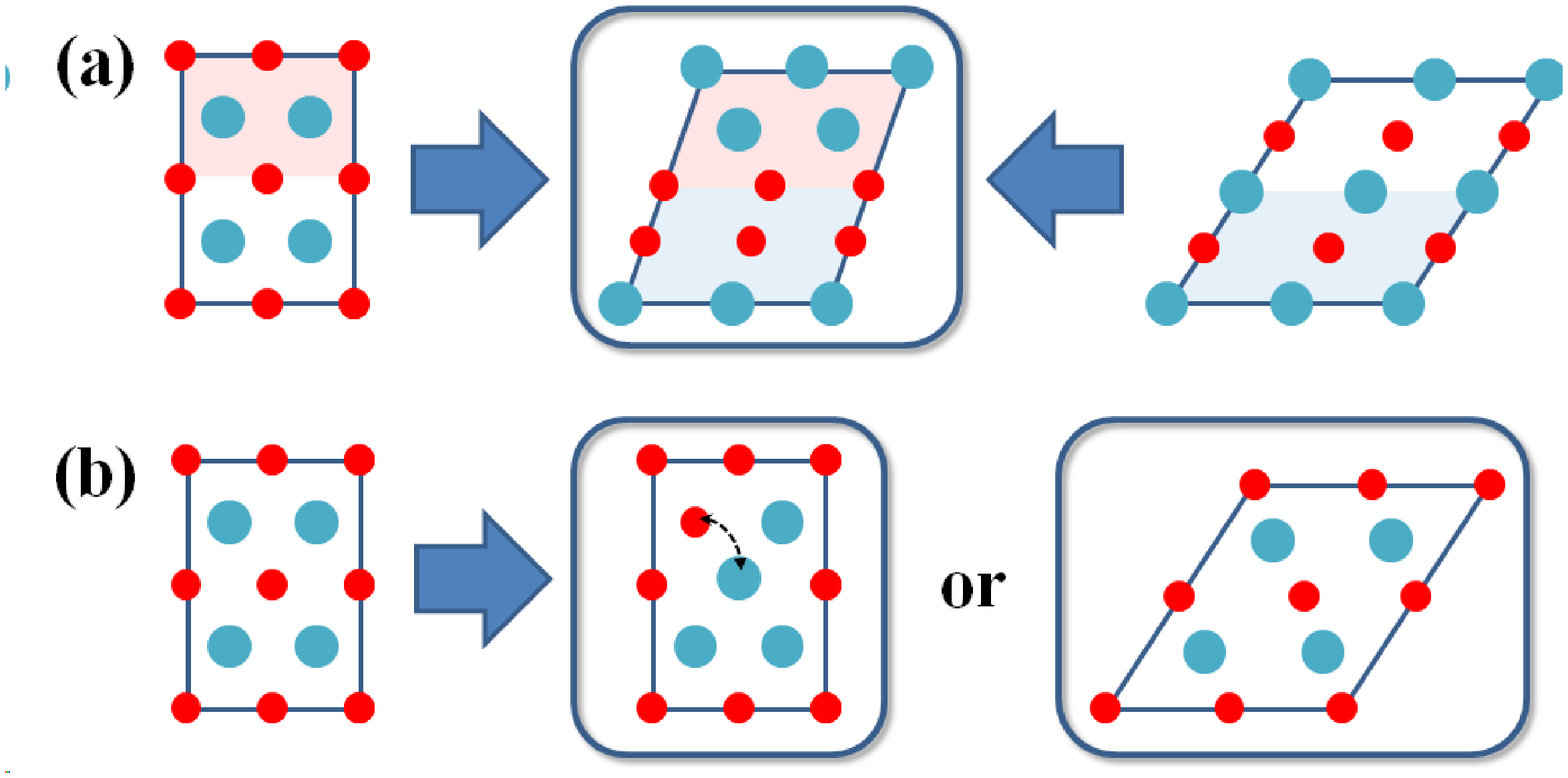}
\caption{\label{Fig-operator-GA} 
(Color online) Schematic view of crystal structure search by the evolutionary operations: (a) mating and (b) mutation.}
\end{figure}

Figure \ref{Fig-flowchart-GA} shows a flowchart at each generation with respect to the structure search by GA. 
For the evolutionary operators, the mating is the operator to create a slab structure from randomly 
selected two individuals (Fig. \ref{Fig-operator-GA} (a)) and the mutation is the operator to give a lattice distortion or an atomic permutation for randomly selected an individual (Fig. \ref{Fig-operator-GA} (b))~\cite{Oganov06}. 
All the structures generated randomly or created by the mating and mutation are optimized at a given pressure, and the enthalpies, \textit{i.e.} $H = E + PV$, are calculated using total energy $E$, pressure $P$, and volume $V$. 
Then, the population for the next generation is constructed by inheriting the few elite structures with especially lower enthalpy at the previous generation, ranking all the structures according to the enthalpy, and eliminating few inferior structures with especially higher enthalpy from the population. 
Once the stable structures are obtained, physical and chemical property data are collected using first-principles calculations (the stage (ii) in Fig. \ref{Fig-MI}). The calculated results are aggregated in the database. 

\subsection{GP for predictor development}

\begin{figure}
\includegraphics[width=8.5cm]{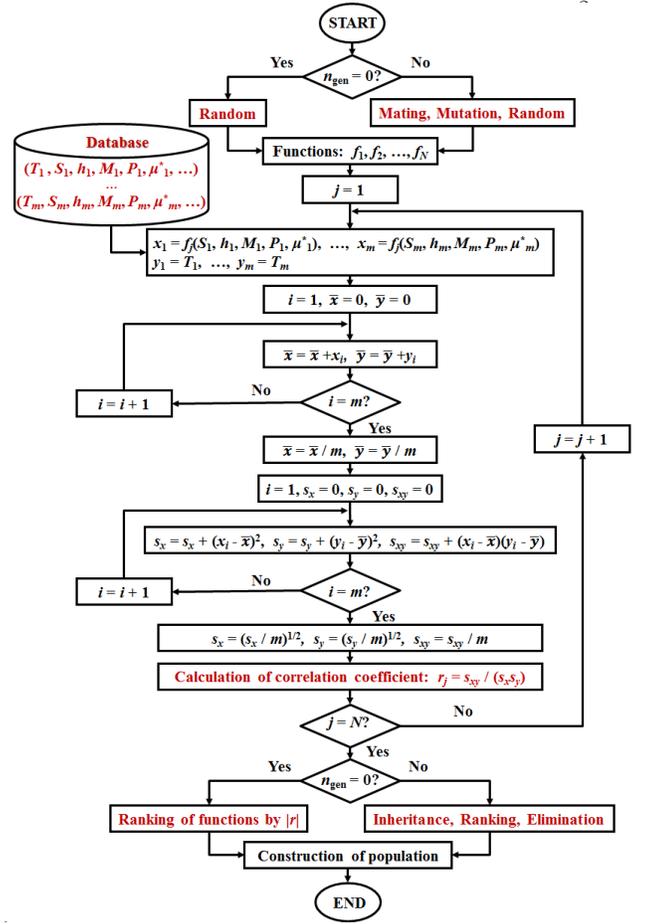}
\caption{\label{Fig-flowchart-GP} 
(Color online) Flowchart at each generation with respect to functionality predictor development by GP. The case of the superconductivity is shown as an example of the functionality (see Sec. \ref{Application_sub1}).}
\end{figure}

\begin{figure}
\includegraphics[width=8.5cm]{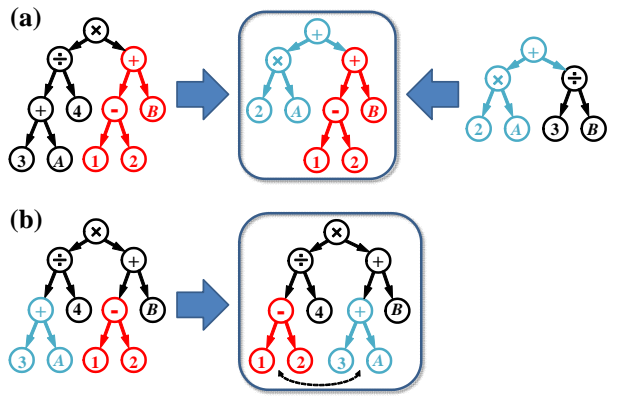}
\caption{\label{Fig-operator-GP} 
(Color online) Schematic view of mathematical function development by evolutionary operations: (a) mating and (b) mutation.}
\end{figure}

Figure \ref{Fig-flowchart-GP} shows a flowchart at each generation with respect to the development of the predictor by GP. 
GP is an expansion of GA where individuals are represented as computer programs or mathematical functions using tree structures as shown in Fig. \ref{Fig-operator-GP}. 
The tree structure consists of branch and leaf nodes, and in our GP search, a branch node depicts an element from arithmetic operators ($+$, $-$, $\times$, $\div$) and a leaf node depicts an element from constants (1-10) and variables to develop mathematical functions. 
For example, the left tree of Fig. \ref{Fig-operator-GP} (a) represents the function of 
$f(A, B) = \{(3 + A) \div 4\} \times \{(1-2) + B\}$, which is written as $f(A, B) = \times\,\div\,+\,3\,A\,4\,+\,-\,1\,2\,B$ in the Polish notation to deal with arithmetic expression in programming language. 
The mating is the operator to create new tree structures 
by exchanging branches between two individuals selected randomly (Fig. \ref{Fig-operator-GP} (a)) and 
the mutation is the operator to exchange branches within an individual (Fig. \ref{Fig-operator-GP} (b)). 
The individuals are ranked according to the strength of the correlation between the value obtained by substituting the datasets into the function (evaluation value: $x$) and physical property data which we will investigate ($y$). 
The correlation coefficient ($r$) between $x$ and $y$ is calculated by $r = s_{xy} / s_{x}s_{y}$, using the data included in the database. 
The parameters $s_{x}$, $s_{y}$, and $s_{xy}$ are a standard deviation of data $x$, 
that of $y$, and a covariance between $x$ and $y$, respectively. They are calculated as follows: 
\begin{equation}
\label{SD_x}
s_{x} = \left( \frac{1}{m} \sum_{i=1}^{m} (x_{i} - \overline{x})^{2} \right)^{1/2},  
\end{equation}
\begin{equation}
\label{SD_y}
s_{y} = \left( \frac{1}{m} \sum_{i=1}^{m} (y_{i} - \overline{y})^{2} \right)^{1/2},  
\end{equation}
and
\begin{equation}
\label{SD_x}
s_{xy} = \frac{1}{m} \sum_{i=1}^{m} (x_{i} - \overline{x})(y_{i} - \overline{y}),  
\end{equation}
where $m$ is the number of the datasets used for the calculation, and 
$\overline{x}$ ($\overline{y}$) is the average value of $x$ ($y$). 
Then, similarly to the case of the structure search, 
a high-quality function is developed by repeatedly performing the evolution, \textit{i.e.} inheriting, ranking according to the absolute value of $r$ ($|r|$), and eliminating. 

For a number of high-quality functions obtained by several GP searches, the predictive ability of the function is checked by a $k$-fold cross-validation~\cite{Stone1974-crossvalidation}. The datasets are randomly divided into $k$ subsets, 
and a single subset is retained as the validation data for testing the functions and the remaining $k-1$ subsets are used for training the functions. 
In this study, we used $R = |r_{ts}|$ for the evaluation of the predictive ability, where $r_{ts}$ represents the correlation coefficient calculated from the testing data and takes the value from 0 to 1. 
If the function developed by GP gives the strong correlation not only for the training data but also the testing data, then $R$ shows a large value. This process is repeated $k$ times by rotation and the values of $R$ are averaged ($\overline{R}$). 
The function with the largest $\overline{R}$ is adopted as the predictor of the physical property which we will investigate. 
At the stage (iv), potential candidates are selected by the regression analysis using the predictor. 

\section{\label{Application}Application to search for superconducting hydrogen compounds}

We developed the calculation codes for the above GA and GP searches and applied our MI method to search for superconducting hydrogen compounds. 
In 2015, high-temperature superconductivity was discovered in hydrogen sulfide (\ce{H_{2}S}) under high pressure and the superconducting critical temperature $T_{\text{c}}$ reaches 203\,K at pressure of 155\,GPa~\cite{Drozdov2015}. 
In 2018-2019, compressed lanthanum hydrides broke the record and 
the superconductivity was observed at 250\,K at 170\,GPa by Drozdov \textit{et al.}~\cite{Drozdov2019-LaHx} and 
260\,K at 190\,GPa by Somayazulu \textit{et al.}~\cite{Somayazulu2019}. 
Therefore, other hydrogen compounds also have a potential to become similar high-$T_{\text{c}}$ superconductors under some conditions. However, 
there are a huge number of combinations for hydrogen compounds in multicomponent system. 
For example, there exist 13572 combinations for ternary hydrogen compounds formed by all elements with the atomic numbers from 2 to 118, and the number is much further increased by taking stoichiometry and crystal structure into account.  Therefore, the MI approach is of great help to discover potential candidates with the high-$T_{\text{c}}$ superconductivity. 

\subsection{Collection of superconductivity data: stage (ii) in MI cycle}

\begin{figure}
\includegraphics[width=8.5cm]{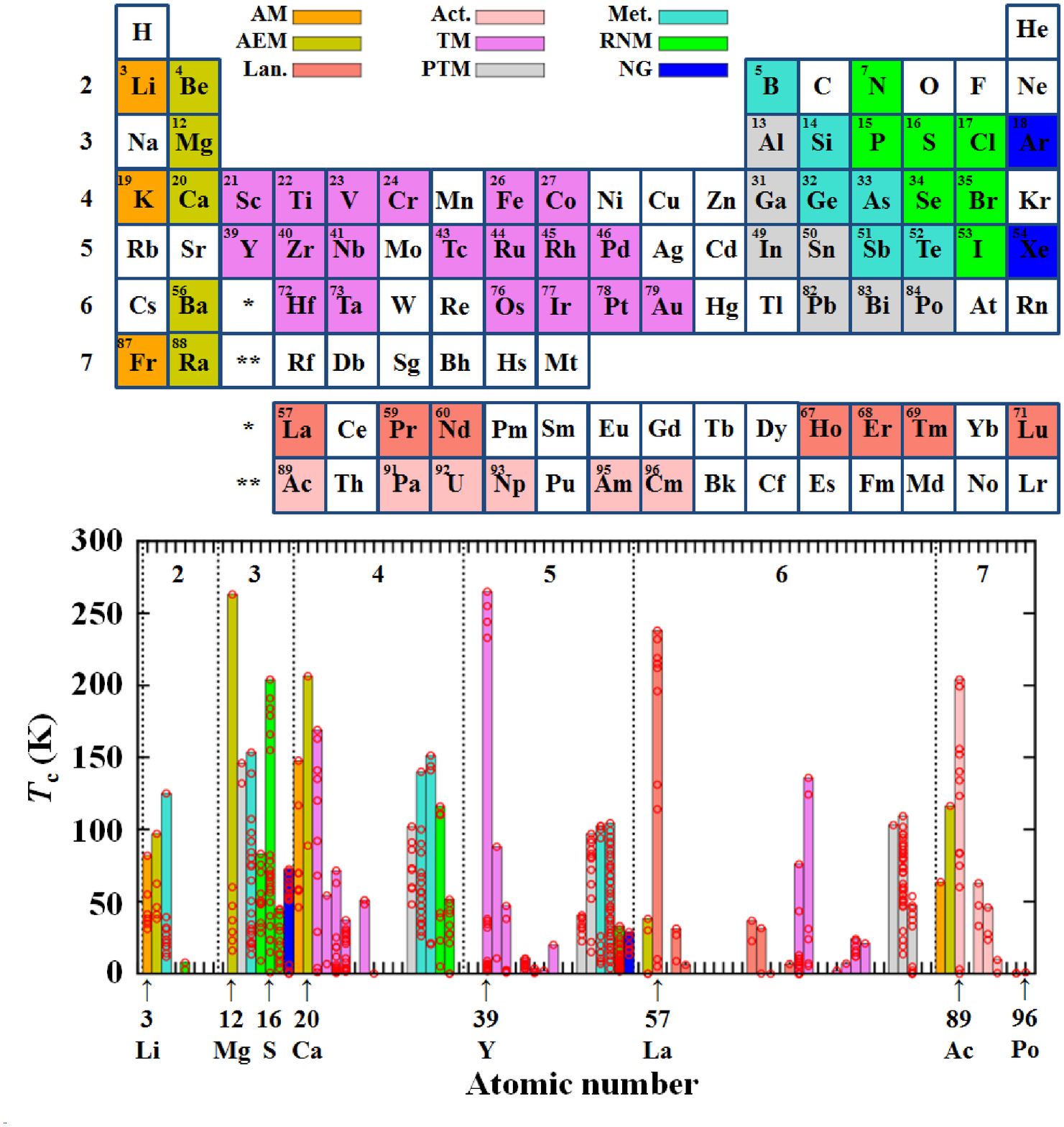}
\caption{\label{Fig-Tcdata} 
(Color online) Elements forming the binary hydrogen compounds included in the database and all the $T_{\text{c}}$ data. The elements are divided into nine groups using different colors: alkali metal (AM), alkaline earth metal (AEM), lanthanide (Lan.), actinide (Act.), transition metal (TM), post-transition metal (PTM), metalloid (Met.), reactive nonmetal (RNM), and noble gas (NG).}
\end{figure}

\begin{figure}
\includegraphics[width=8.2cm]{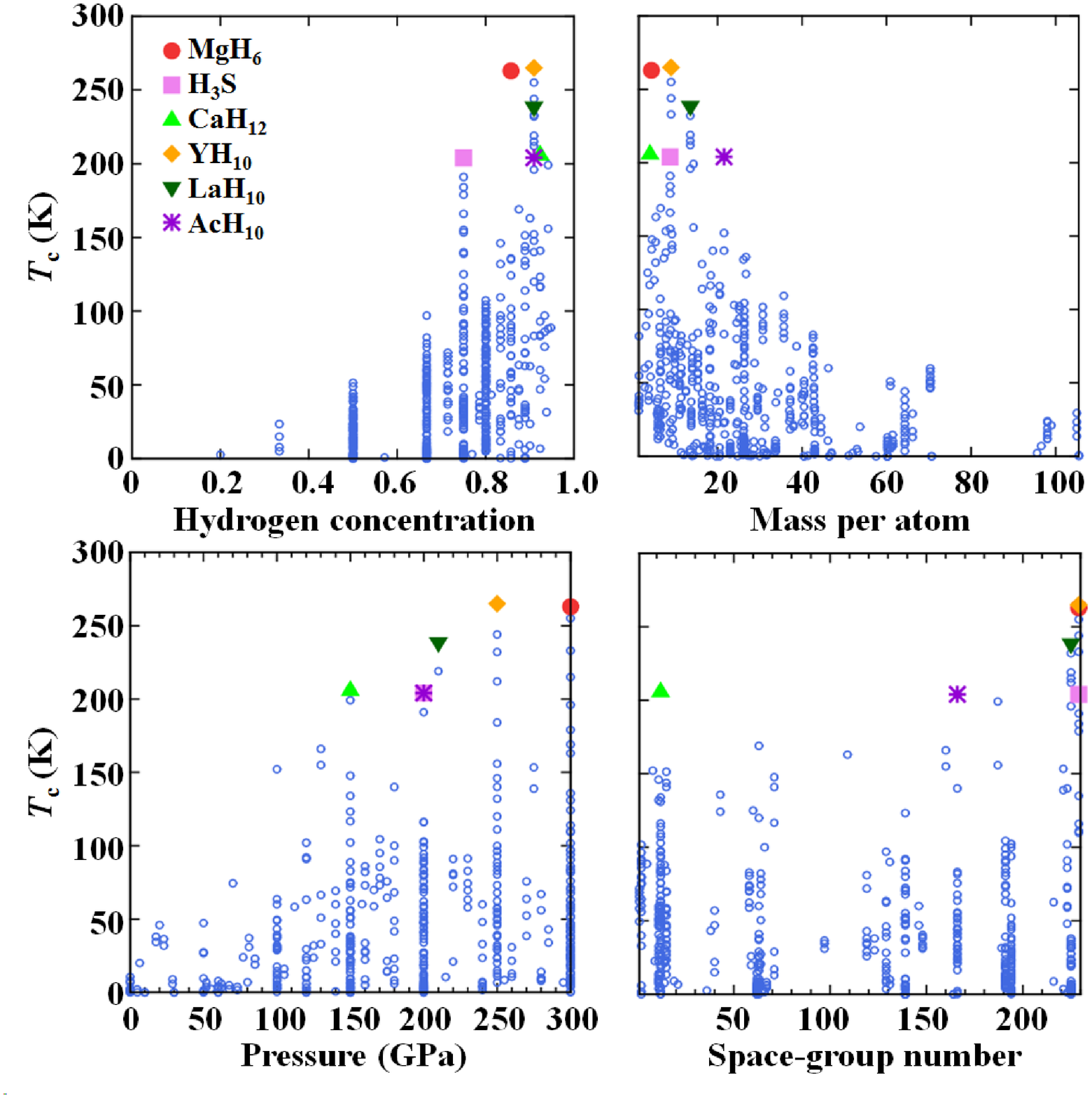}
\caption{\label{Fig-database} 
(Color online) Relationships between the superconducting $T_{\text{c}}$ and hydrogen concentration (top left), $T_{\text{c}}$ and mass per atom (top right), $T_{\text{c}}$ and pressure (bottom left), and $T_{\text{c}}$ and space-group number (bottom right) for superconducting binary hydrogen compounds predicted by first-principles calculations.}
\end{figure}

First we developed the database by collecting the first-principles calculation data with respect to chemical composition, crystal structure, and superconducting property of binary hydrogen compounds under high pressure condition from literature~\cite{Chen2008,Kim2009,Tse2009-BaH2,Jin2010-disilane,Duan2010,Gao2010,Kim2011,Gao2011,Abe2011,Zhang2012,Zhou2012,
Lonie2013,Gao2013,Hu2013,Hooper2013,Abe2013,Duan2014-SciRep,Chen2014,Errea2014,Yu2014,Li2014,Xie2014-Li,Wang2014-Be,Chen2015,Duan2015-iodine,Shamp2015,Zhang2015,Zhang2015-SiH4,Errea2015,Hou2015,Yu2015,Zhang2015-selenium,Feng2015,Yan2015-Xe,Cheng2015-Pb,Liu2015-arXiv,Liu2015-indium,Liu2015-osmium,Liu2015-Hf,Ma2015-arXiv,Shamp2016,Liu2016-PH3,Qian2016-arXiv,Esfahani2016,Ishikawa2016-SciRep,Li2016-technetium,Zhong2016-tellurium,Fu2016-pnictogen,Li2016,Liu2016-Ru,Majumdar2017,Liu2017-LaH10,Esfahani2017-arXiv,Zeng2017,Zhuang2017,Ishikawa2017-ArHx,Li2017-vanadium,Li2017-zirconium,Semenok2018-arXiv,Semenok2018,Shanavas2018,Wang2018,Ye2018-scandium}. 
The database includes 497 datasets for the compounds with 62 elements colored in the periodic table of Fig. \ref{Fig-Tcdata}. 
We used only the $T_{\text{c}}$ data obtained by the Allen-Dynes formula~\cite{Allen-Dynes} in this study, 
which are all plotted in the lower panel of Fig. \ref{Fig-Tcdata}. 
The highest $T_{\text{c}}$ value for each binary system is represented as a histogram. 
The superconductivity of $T_{\text{c}} \geq 200$\,K is predicted in the following compounds: $T_{\text{c}} = 265$\,K at 250\,GPa in \ce{YH_{10}}~\cite{Liu2017-LaH10}, 263\,K at 300\,GPa in \ce{MgH_{6}}~\cite{Feng2015}, 238\,K at 210\,GPa in \ce{LaH_{10}}~\cite{Liu2017-LaH10}, 206\,K at 150\,GPa in \ce{CaH_{12}}~\cite{Semenok2018-arXiv}, 204\,K at 200\,GPa in \ce{AcH_{10}}~\cite{Semenok2018}, and 204\,K at 200\,GPa in \ce{H_{3}S}~\cite{Duan2014-SciRep}. 
Figure \ref{Fig-database} shows the relationships between $T_{\text{c}}$ and hydrogen concentration, mass per atom, pressure, and space group number for all the data included in the database. 
Although large hydrogen concentration, light mass, and high pressure favor the increase of the superconductivity, they are not a necessary and sufficient condition for high-$T_{\text{c}}$ superconductivity. Further, space group number, \textit{i.e.} crystal symmetry, shows a weak correlation with $T_{\text{c}}$. The data suggests that the search for hydrogen compounds with high $T_{\text{c}}$ is seemingly simple, but it is actually difficult just to investigate these parameters independently. 

\subsection{\label{Application_sub1}Development of superconductivity predictor: stage (iii) in MI cycle}

Next we developed the superconductivity predictor by GP using the datasets included in the database. 
The datasets were divided into ten subsets for the cross-validation, \textit{i.e.} $k = 10$, and 90\% of the data for training and 10\% for testing. 
We used five variables, space group ($S$), hydrogen concentration ($h$), mass per atom ($M$), pressure ($P$), and the effective screened Coulomb repulsion constant in the Allen-Dynes formula~\cite{Allen-Dynes} ($\mu^{*}$), to obtain the superconductivity evaluation value $x$; $x = f(S, h, M, P, \mu^{*})$ (see Fig. \ref{Fig-flowchart-GA}). 
The functions are evolved for 5000 generations, 
based on the correlation to the superconducting $T_{\text{c}}$ ($T$); $y = T$. 
The number of the individuals were set at 20, in which 8 functions are created by the mating, 8 by the mutation, 2 are randomly generated, and 2 are inherited. 
Then the 10-fold cross-validation was carried out and the $\overline{R}$ value was calculated. 
We performed this process 50 times in parallel and adopted the function with the largest $\overline{R}$ as the superconductivity predictor. 
Figure \ref{Fig-GPresult} shows the correlation between $T_{\text{c}}$ and $x$ obtained by the predictor. $|r_{tr}|$ and $|r_{ts}|$ are 0.78 and 0.81, respectively, where $r_{tr}$ represents the correlation coefficient calculated from the training data and takes the value from 0 to 1. 
The compounds taking small $x$ values are potential candidates for the high-$T_{\text{c}}$ superconductivity. 
See Ref. 80 with respect to the details of the predictor, 
in which the function is represented as the Polish notation. 

\subsection{\label{Application_sub3}Search for potential candidates: stage (iv) in MI cycle}

\begin{figure}
\includegraphics[width=7.5cm]{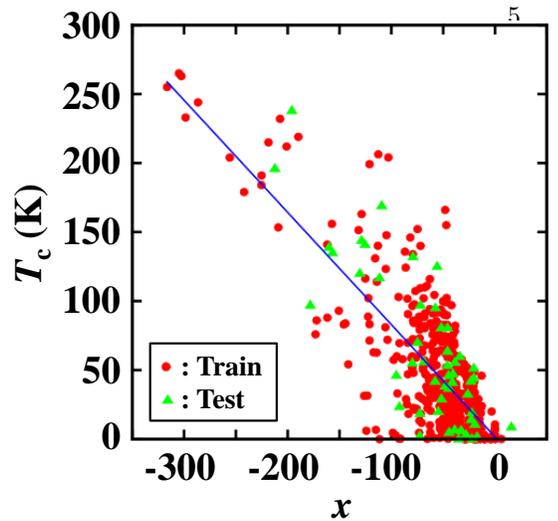}
\caption{\label{Fig-GPresult} 
(Color online) Correlation between $T_{\text{c}}$ and the superconductivity evaluation value $x$ calculated by the predictor. The training and testing data are represented as circle and triangle, respectively. }
\end{figure}

We searched for the potential candidates in ternary hydrogen compounds using this predictor. 
First, we created 497 datasets for ternary hydrogen compounds using all the datasets for the binary ones included in our database. 
That is to say, if in a binary compound the element paired with hydrogen has the atomic number of $Z$, then the corresponding ternary compound is created by replacing it with the elements having $Z-1$ and $Z+1$ as follows: 
\ce{H_{6}P(\textit{Z} = 15)Cl(\textit{Z} = 17)} for \ce{H_{3}S(\textit{Z} = 16)}, \ce{Ba(\textit{Z} = 56)Ce(\textit{Z} = 58)H_{20}} for \ce{La(\textit{Z} = 57)H_{10}}, \textit{etc.}. 
The values of $S$, $h$, $P$, and $\mu^{*}$ were 
matched to those of the corresponding binary data, in other word, only $M$ were varied. 
Then, we calculated the $x$ values of the hypothetical ternary compounds using the created datasets. 
Table \ref{results-ternary} lists a part of the results on $x$ and $T_{\text{c}}$ for the hypothetical ternary compounds. We estimated the $T_{\text{c}}$ values of the ternary compounds ($T_{\text{c}}^{\text{t}}$) using the relationship of $T_{\text{c}}^{\text{t}} = -0.8130 (x^{\text{t}}-x^{\text{b}}) + T_{\text{c}}^{\text{b}}$, where $x^{\text{t}}$ ($x^{\text{b}}$) is the evaluation value of the ternary (binary) compound and $T_{\text{c}}^{\text{b}}$ is the $T_{\text{c}}$ value of the binary compounds obtained by the Allen-Dynes formula~\cite{Allen-Dynes}, included in the database. The slope of $-0.8130$ 
is taken from a simple linear regression model, $y = -0.8130x + 1.66$, 
which is obtained by a least-squares fitting of the data (see the line in Fig. \ref{Fig-GPresult}). 
In the table, the results are classified into (a) the group 1-3 elements and (b) the group 13-16 elements. 
For the group 1-3 elements, in which polyhydrides with hydrogen cages are formed and $T_{\text{c}}$ exceeds 200\,K, the superconductivity is slightly suppressed for \ce{YH_{10}} and \ce{MgH_{6}} and is less affected for \ce{LaH_{10}} and \ce{AcH_{10}} by changing to the hypothetical ternary compounds. 
Only \ce{CaH_{12}} shows a slight enhancement of the superconductivity. 
For the group 13-16 elements, in which covalent hydrides are formed and $T_{\text{c}}$ shows the values from 140 to 200\,K, the superconductivity is largely suppressed for \ce{H_{3}S} and \ce{SiH_{3}}, whereas it is less affected for \ce{AsH_{8}}, \ce{AlH_{5}}, and \ce{GeH_{3}}.  
\begin{table}
\caption{\label{results-ternary}
Superconductivity of the hypothetical ternary hydrogen compounds. The $T_{\text{c}}$ values of the ternary compounds are estimated from the amount of change in the superconductivity evaluation value $x$.}
\begin{ruledtabular}
\begin{tabular}{lcc}
compounds & $x$ & $T_{\text{c}}$ (K) \\
$(S, h, P, \mu^{*})$ & ($M$)  & \\ \hline
(a) group 1-3 & & \\
\ce{YH_{10}} $\rightarrow$ \ce{SrZrH_{20}} & -304.9 $\rightarrow$ -292.7 & 265~\cite{Liu2017-LaH10} $\rightarrow$ 255 \\
(229, 0.9091, 250, 0.10) & (9.00 $\rightarrow$ 9.05) & \\
\ce{MgH_{6}} $\rightarrow$ \ce{NaAlH_{12}} & -302.6 $\rightarrow$ -287.5 & 263~\cite{Feng2015} $\rightarrow$ 251\\
(229, 0.8571, 300, 0.12) & (4.34 $\rightarrow$ 4.43) & \\
\ce{LaH_{10}} $\rightarrow$ \ce{BaCeH_{20}} & -196.1 $\rightarrow$ -196.4 & 238~\cite{Liu2017-LaH10} $\rightarrow$ 238\\
(225, 0.9091, 210, 0.10) & (13.54 $\rightarrow$ 13.53) & \\
\ce{CaH_{12}} $\rightarrow$ \ce{KScH_{24}} & -112.6 $\rightarrow$ -113.7 & 206~\cite{Semenok2018-arXiv} $\rightarrow$ 207\\
(12, 0.9231, 150, 0.10) & (4.01 $\rightarrow$ 4.16) & \\
\ce{AcH_{10}} $\rightarrow$ \ce{RaThH_{20}} & -102.9 $\rightarrow$ -102.6 & 204~\cite{Semenok2018} $\rightarrow$ 204\\ 
(166, 0.9091, 200, 0.10) & (21.55 $\rightarrow$ 21.73) & \\ \hline
(b) group 13-16 & & \\
\ce{H_{3}S} $\rightarrow$ \ce{H_{6}PCl} & -255.9 $\rightarrow$ -152.2 & 204~\cite{Duan2014-SciRep} $\rightarrow$ 120\\
(229, 0.7500, 200, 0.10) & (8.77 $\rightarrow$ 9.06) & \\
\ce{SiH_{3}} $\rightarrow$ \ce{AlPH_{6}} & -209.1 $\rightarrow$ -50.2 & 153~\cite{Jin2010-disilane} $\rightarrow$ 24\\
(221, 0.7500, 275, 0.10) & (7.78 $\rightarrow$ 8.00) & \\
\ce{AsH_{8}} $\rightarrow$ \ce{GeSeH_{16}} & -131.6 $\rightarrow$ -131.6 & 151~\cite{Fu2016-pnictogen} $\rightarrow$ 151\\
(15, 0.8889, 450, 0.10) & (9.22 $\rightarrow$ 9.32) & \\
\ce{AlH_{5}} $\rightarrow$ \ce{MgSiH_{10}} & -81.7 $\rightarrow$ -81.6 & 146~\cite{Hou2015} $\rightarrow$ 146\\
(11, 0.8333, 250, 0.10) & (5.34 $\rightarrow$ 5.21) & \\
\ce{GeH_{3}} $\rightarrow$ \ce{GaAsH_{6}} & -71.9 $\rightarrow$ -72.1 & 140~\cite{Abe2013} $\rightarrow$ 140\\
(223, 0.7500, 180, 0.13) & (18.91 $\rightarrow$ 18.84) & \\
\end{tabular}
\end{ruledtabular}
\end{table}

\subsection{Validation of superconductivity from first principles: stages (i) and (ii) in MI cycle}

\subsubsection{\ce{KScH_{12}}}

We verified the superconductivity of one of the potential candidates, \ce{KScH_{12}}, which is a hypothetical ternary compound approximated to \ce{CaH_{6}} and is included in the group 1-3 elements. \ce{CaH_{6}} is predicted to form the hydrogen cage structure under high pressure and show $T_{\text{c}}$ of 220-235\,K at 150\,GPa, which is obtained by numerically solving the Eliashberg equations~\cite{Wang2012}. Therefore, \ce{KScH_{12}} is  expected to show similar crystal structure and superconductivity under high pressure. 

\begin{figure}
\includegraphics[width=8.2cm]{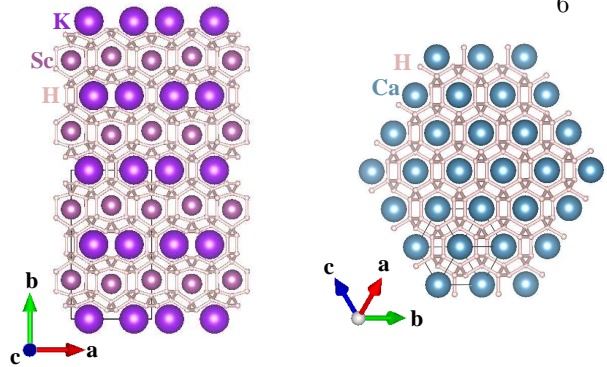}
\caption{\label{Fig-KScH12} 
(Color online) (Left) Crystal structure of \ce{KScH_{12}}, predicted by GA. The space group is a monoclinic $C2/m$ (No. 12) with $a = 5.3726$\AA, $b/a = 1.7795$, $c/a = 0.5131$, $\beta = 106.6^{\circ}$, K: $4i$ (0.77998, 0, 0.47704), Sc: $4g$ (0, 0.26336, 0), H1: $8j$ (0.74515, 0.37079, 0.01997), H2: $8j$ (-0.05716, 0.43888, 0.14288), H3: $8j$ (0.57041, 0.41013, -0.05075), H4: $8j$ (0.81603, 0.18680, 0.40624), H5: $8j$ (0.68641, 0.18066, 0.59388), H6: $4h$ (0, 0.38665, 0.5), and H7: $4h$ (0, 0.84338, 0.5) at 300\,GPa. (Right) Crystal structure of $Im$-3$m$ \ce{CaH_{6}} reported earlier~\cite{Wang2012}. Crystal structures were drawn with VESTA~\cite{VESTA}.}
\end{figure}

The variation of $T_{\text{c}}$ shown in Table \ref{results-ternary} is estimated under the hypothesis that the space group $S$ is invariant between binary and ternary compounds. 
However, $S$ for the most stable structure of the ternary compound is considered to be different from that of the binary one.   
Therefore, first we searched for the most stable structure of \ce{KScH_{12}} by combining our GA code with the Quantum ESPRESSO (QE) code~\cite{QE} and applying it to \ce{KScH_{12}}. 
In our GA search, 8 structures are created by 
``mating'', 6 ``mutation (distortion)'', and 6 ``mutation (permutation)'' in each generation. 
We performed the structure search at pressures of 200 and 300\,GPa, 
using calculation cells including 2 formula units. 
The generalized gradient approximation by 
Perdew, Burke and Ernzerhof~\cite{PBE} was used for the exchange-correlation functional, and 
the Rabe-Rappe-Kaxiras-Joannopoulos ultrasoft pseudopotential~\cite{RRKJ90} was employed. 
The $k$-space integration over the Brillouin zone (BZ) was carried out 
on a 4 $\times$ 4 $\times$ 4 grid, and the energy cutoff was set at 80\,Ry for the wave function and 640\,Ry for the charge density. 
After we obtained the stable structures, we improved the calculation accuracy
by increasing the number of k-points and checked the stability. 
As the results, we obtained a monoclinic $C2/m$ (No. 12) structure at 300\,GPa, where it shows no phonon instability. 
The structure takes a modulated hydrogen-cage structure, similar to the structure of \ce{CaH_{6}} (compare the structure between the left and right in Fig. \ref{Fig-KScH12}). 
The evaluation values of $Im$-3$m$ \ce{CaH_{6}} ($S = 229$) and $C2/m$ \ce{KScH_{12}} ($S = 12$) are calculated to be $-179.5$ and $-95.5$ at 300\,GPa, respectively (Table \ref{evaluationvalue-KScH12}). That is to say, $T_{\text{c}}$ decreases by about 68\,K owing to the change from $Im$-3$m$ \ce{CaH_{6}} to $C2/m$ \ce{KScH_{12}} according to the slope of $-0.8130$ mentioned in \ref{Application_sub3}. 
\begin{table}
\caption{\label{evaluationvalue-KScH12}
Variation of superconductivity evaluation value $x$ and superconducting $T_{\text{c}}$ ($\Delta T_{\text{c}}$) due to the change from $Im$-3$m$ \ce{CaH_{6}} to $C2/m$ \ce{KScH_{12}} at 300\,GPa.}
\begin{ruledtabular}
\begin{tabular}{ccc}
\ce{CaH_{6}} $\rightarrow$ \ce{KScH_{12}} & $x$ & $\Delta T_{\text{c}}$ (K) \\ \hline
$Im$-3$m$ $\rightarrow$ $C2/m$ & -179.5 $\rightarrow$ -95.5 & $-68$\\
\end{tabular}
\end{ruledtabular}
\end{table}

Next, we investigated the superconductivity of $C2/m$ \ce{KScH_{12}} using first-principles calculations. 
$T_{\text{c}}$ was calculated using the Allen-Dynes formula~\cite{Allen-Dynes}, 
\begin{equation}
\label{AllenDynes}
T_{\text{c}}=\frac{\omega_{\log}}{1.2}
 \exp \left[ - \frac{ 1.04(1+\lambda )}
{ \lambda-\mu^{\ast}(1+0.62\lambda) } \right].
\end{equation}
In Eq. \ref{AllenDynes}, the parameters of electron-phonon coupling constant $\lambda$ and 
logarithmic-averaged phonon frequency $\omega_{\log}$  
represent a set of characters for the phonon-mediated superconductivity. 
To obtain these parameters, we performed the phonon calculations implemented in the QE code. We used an 8 $\times$ 16 $\times$ 8 $k$-point grid for the electron-phonon calculation, and 4 $\times$ 8 $\times$ 4 $k$-point and 4 $\times$ 4 $\times$ 4 $q$-point grids for the dynamical matrix calculation. 
We also calculated $T_{\text{c}}$ of $Im$-3$m$ \ce{CaH_{6}} using a 24 $\times$ 24 $\times$ 24 $k$-point, and 12 $\times$ 12 $\times$ 12 $k$-point and 4 $\times$ 4 $\times$ 4 $q$-point grids. 
The effective screened Coulomb repulsion constant $\mu^{*}$ 
was assumed to be 0.13, which has been considered to be a reasonable value 
for hydrides. 
The results are listed in Table \ref{KScH12-superconductivity}. 
$T_{\text{c}}$ of $Im$-3$m$ \ce{CaH_{6}} at 150\,GPa is 172\,K, which is lower by about 50\,K than that obtained by numerically solving the Eliashberg equations reported earlier~\cite{Wang2012}. 
$T_{\text{c}}$ of $C2/m$ \ce{KScH_{12}} shows 122\,K at 300\,GPa, 
which is decreased by 38\,K owing to the change from $Im$-3$m$ \ce{CaH_{6}}. 
This result indicates that the decrease of $T_{\text{c}}$ is qualitatively consistent with that of estimated from the variation of the evaluation value $x$ shown in Table \ref{evaluationvalue-KScH12}. 
\begin{table}
\caption{\label{KScH12-superconductivity}
Superconductivity of \ce{CaH_{6}} and \ce{KScH_{12}}, obtained by the first-principles calculations and the Allen-Dynes formula. $T_{\text{c}}$ was calculated using $\mu^{*} = 0.13$.
}
\begin{ruledtabular}
\begin{tabular}{cccccc}
          &  space group & $P$ (GPa) & $\lambda$ & $\omega_{\text{ln}}$ & $T_{\text{c}}$ (K)\\ \hline
\ce{CaH_{6}}                  & $Im$-3$m$ (No.229) &  150         & 2.55       & 1091                    & 172\\
                  &  &  300         & 1.66       & 1388                    & 160\\
 \hline
\ce{KScH_{12}} & $C2/m$ (No.12) & 300          & 1.54       & 1139                    & 122\\
\end{tabular}
\end{ruledtabular}
\end{table}

\subsubsection{\ce{GaAsH_{6}}}

\begin{figure}
\includegraphics[width=8.2cm]{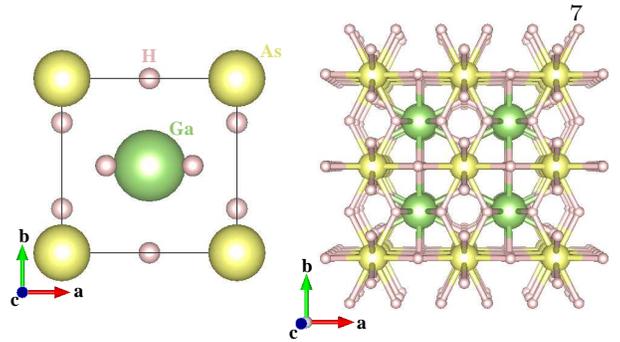}
\caption{\label{Fig-GaAsH6} 
(Color online) Crystal structure of \ce{GaAsH_{6}}, predicted by GA. The space group is a cubic $Pm$-3 (No. 200) with $a = 3.1050$\AA, Ga: $1b$ (0.5, 0.5, 0.5), As: $1a$ (0, 0, 0), and H: $6g$ (0.74736, 0.5, 0). Crystal structure was drawn with VESTA~\cite{VESTA}.}
\end{figure}

We verified the superconductivity of another candidate, \ce{GaAsH_{6}}, included in the group 13-16 elements. Gallium arsenide (\ce{GaAs}) has been well known as a III-V direct semiconductor at ambient pressure, and the hydrogenation of the compound is expected to be achieved under high-pressure conditions, as well as group IV semiconductors. We searched for stable structures at pressures of 50, 100, 200, and 300\,GPa, using calculation cells including 2  formula units. 
The $k$-space integration over the Brillouin zone (BZ) was performed 
on an 8 $\times$ 8 $\times$ 8 grid, and the energy cutoff was set at 80\,Ry for the wave function and 640\,Ry for the charge density. 
As the results, we obtained a cubic $Pm$-3 (No. 200) structure at 180\,GPa (Fig. \ref{Fig-GaAsH6}). Although the structure is similar to the A15 structure with $Pm$-3$n$ (No. 223) of \ce{GeH_{3}} reported earlier~\cite{Abe2013}, the space group is different from that of the A15 structure. 
The evaluation value $x$ of $Pm$-3 \ce{GaAsH_{6}} is calculated to be $-40.0$ at 180\,GPa, 
which gives the $T_{\text{c}}$ variations of $-8$\,K from $Cccm$ \ce{GeH_{3}}, 
$-7$\,K from $P4_{2}/mmc$ \ce{GeH_{3}}, and 
$-26$\,K from $Pm$-3$n$ \ce{GeH_{3}} (Table \ref{evaluationvalue-GaAsH6}). 
That is to say, $T_{\text{c}}$ is slightly varied according to the change from $Cccm$ or $P4_{2}/mmc$ \ce{GeH_{3}} to $Pm$-3 \ce{GaAsH_{6}}, whereas it is more largely decreased according to the change from $Pm$-3$n$ \ce{GeH_{3}}. 
\begin{table}
\caption{\label{evaluationvalue-GaAsH6}
Variation of superconductivity evaluation value $x$  and superconducting $T_{\text{c}}$ ($\Delta T_{\text{c}}$) due to the change from \ce{GeH_{3}} to \ce{GaAsH_{6}} at 180\,GPa. The three datasets of \ce{GeH_{3}}, $Pm$-3$n$ (No. 223) \ce{GeH_{3}}, $Cccm$ (No. 66) \ce{GeH_{3}}, and $P4_{2}/mmc$ (No. 131), are taken from Ref. \cite{Abe2013}.}
\begin{ruledtabular}
\begin{tabular}{ccc}
\ce{GeH_{3}} $\rightarrow$ \ce{GaAsH_{6}} & $x$ & $\Delta T_{\text{c}}$ (K) \\ \hline
$Cccm$ $\rightarrow$ $Pm$-3 & -49.7 $\rightarrow$ -40.0 & $-8$\\
$P4_{2}/mmc$ $\rightarrow$ $Pm$-3 & -48.8 $\rightarrow$ -40.0 & $-7$\\
$Pm$-3$n$ $\rightarrow$ $Pm$-3 & -71.9 $\rightarrow$ -40.0 & $-26$\\
\end{tabular}
\end{ruledtabular}
\end{table}

We investigated the superconductivity of $Pm$-3 \ce{GaAsH_{6}} using first-principles calculations and the Allen-Dynes formula. 
We used a 48 $\times$ 48 $\times$ 48 $k$-point grid for the electron-phonon calculation, and 24 $\times$ 24 $\times$ 24 $k$-point and 6 $\times$ 6 $\times$ 6 $q$-point grids for the dynamical matrix calculation. 
The effective screened Coulomb repulsion constant $\mu^{*}$ 
was assumed to be 0.13. 
The results are listed in Table \ref{GaAsH6-superconductivity}. 
$T_{\text{c}}$ of $Pm$-3 \ce{GaAsH_{6}} shows 98\,K at 180\,GPa, 
which is lower by 42\,K than that of a metastable $Pm$-3$n$ phase of \ce{GeH_{3}} ($T_{\text{c}} = 140$\,K) but is comparable with that of the most stable $Cccm$ phase ($T_{\text{c}} = 100$\,K). 
Thus, the $T_{\text{c}}$ variations are qualitatively consistent with those estimated from the evaluation value $x$. 

These results suggest that the regression analysis based on the superconductivity predictor gives reliable results on potential candidates for the high-$T_{\text{c}}$ superconductivity in the ternary hydrogen compounds approximated to binary ones included in the database. 
\begin{table}
\caption{\label{GaAsH6-superconductivity}
Superconductivity of \ce{GeH_{3}} and \ce{GaAsH_{6}} obtained by the first-principles calculations and the Allen-Dynes formula. The data of \ce{GeH_{3}} is taken from the Ref. \cite{Abe2013}. $T_{\text{c}}$ was calculated using $\mu^{*} = 0.13$.
}
\begin{ruledtabular}
\begin{tabular}{cccccc}
          &  space group & $P$ (GPa) & $\lambda$ & $\omega_{\text{ln}}$ & $T_{\text{c}}$ (K)\\ \hline
\ce{GeH_{3}}                  & $Cccm$ (No.66) &  180         & 1.60       & 793                    & 100\\
                  & $P4_{2}/mmc$ (No.131) &          & 1.56       & 737                    & 90\\
 & $Pm$-3$n$ (No.223) &          & 1.82       & 989                    & 140\\ \hline
\ce{GaAsH_{6}} & $Pm$-3 (No.200) & 180          & 1.57       & 897                    & 98\\
 & & 200          & 1.43       & 964                    & 96\\
& & 300          & 1.57       & 635                    & 69\\
\end{tabular}
\end{ruledtabular}
\end{table}

\section{\label{CON}Discussion and conclusion}

We developed the MI method consisting of the four stages, which is based on EA: (i) the stable structures are determined using GA, (ii) the physical and chemical property data are collected by performing first-principles calculations for the stable structures, (iii) the functionality predictor is developed from the datasets in the database using GP, and (iv) potential candidates are discovered by the regression analysis. Turning of the cycle enables to accelerate the search for novel functional materials. 

We applied the MI method to search for the superconductivity in hydrogen compounds. First, we developed the database on the superconductivity of the binary hydrogen compounds by collecting the data from literature. Then, we developed the superconductivity predictor and explored the superconductivity in the hypothetical ternary hydrogen compounds, approximated to the binary compounds included in the database. For the group 1-3 elements, the superconductivity is less affected by changing to the ternary compounds, and high-$T_{\text{c}}$ superconductivity is expected in the hypothetical ternary compounds, as is the case with the binary ones. For the group 13-16 elements, the superconductivity is largely suppressed in the ternary compounds approximated to \ce{H_{3}S} and \ce{SiH_{3}}, whereas those approximated to \ce{AsH_{8}}, \ce{AlH_{5}}, and \ce{GeH_{3}} are less affected by the changing.  
We actually verified the superconductivity of \ce{KScH_{12}} created from \ce{CaH_{6}} and \ce{GaAsH_{6}} created from \ce{GeH_{3}} using the first-principles calculations. For \ce{KScH_{12}}, a modulated hydrogen-cage structure with a space group of $C2/m$ is obtained at 300\,GPa by GA, and $T_{\text{c}}$ shows 122\,K for $\mu^{*} = 0.13$. 
For \ce{GaAsH_{6}}, we predicted a cubic $Pm$-3 structure and obtained $T_{\text{c}}$ of 98\,K at 180\,GPa. 
These $T_{\text{c}}$ values are qualitatively consistent with those estimated from the variations of the superconductivity evaluation values. These results suggest that the regression analysis based on the superconductivity predictor is effective for the discovery of potential candidates in the ternary hydrogen compounds. 

The prediction ability of the superconductivity in ternary hydrogen compounds is expected to be further improved by aggregating the superconductivity data of \ce{KScH_{12}}, \ce{GaAsH_{6}}, and other ternary compounds into the database and redeveloping the predictor by GP. 
At present, the superconductivity of $T_{\text{c}} \geq 200$\,K is found only in the pressure above 150\,GPa. Therefore, the achievement of more than 200\,K superconductivity at low pressure is also significantly important, and MI approaches could be effective for the research.

\begin{acknowledgments}
This work was supported by JSPS KAKENHI under Grant-in-Aid for Specially Promoted Research (26000006), Scientific Research (C) (17K05541), and Scientific Research (S) (16H06345), Asahi Glass Foundation, Yamada Science Foundation, and MEXT as ``Exploratory Challenge on Post-K computerh (Frontiers of Basic Science: Challenging the Limits).
\end{acknowledgments}
\bibliography{References}

\end{document}